\documentclass[aps,pra,twocolumn,groupedaddress,superscriptaddress,bibnotes,amsfonts,citeautoscript,a4paper]{revtex4-1}
\usepackage{latexsym}
\usepackage{amsmath}
\usepackage{amssymb}
\usepackage{graphicx}
\usepackage{epstopdf}
\usepackage[T1]{fontenc}
\usepackage[open]{bookmark}
\usepackage{hyperref}
\hypersetup{colorlinks=true,allcolors=blue}
\usepackage{orcidlink}
\usepackage{epsfig}
\usepackage{epstopdf}
\usepackage{soul}
\usepackage{float}
\usepackage{ragged2e}
\usepackage{amsfonts}
\usepackage{color}
\usepackage{dsfont}
\usepackage[percent]{overpic}
\usepackage{hyperref}
\usepackage{lineno}
\usepackage[normalem]{ulem}
\usepackage{mathrsfs}

\begin{document}

\title{Self-similar plasmonic nanolenses: mesoscopic ensemble averaging and chiral light--matter interactions}

\author{Nikolaos Kyvelos\,\orcidlink{0000-0002-5164-1479}}
\email{niky@mci.sdu.dk}
\affiliation{POLIMA---Center for Polariton-driven Light--Matter Interactions, University of Southern Denmark, Campusvej 55, DK-5230 Odense M, Denmark}
\author{N. Asger Mortensen\,\orcidlink{0000-0001-7936-6264}}
\affiliation{POLIMA---Center for Polariton-driven Light--Matter Interactions, University of Southern Denmark, Campusvej 55, DK-5230 Odense M, Denmark}
\affiliation{Danish Institute for Advanced Study, University of Southern Denmark,
Campusvej 55, DK-5230 Odense M, Denmark}
\author{Xuezhi Zheng\,\orcidlink{0000-0002-7832-7527}}
\affiliation{Department of Electrical Engineering (ESAT-WaveCore), KU Leuven, B-3001 Leuven, Belgium}
\author{Christos Tserkezis\,\orcidlink{0000-0002-2075-9036}}
\email{ct@mci.sdu.dk}
\affiliation{POLIMA---Center for Polariton-driven Light--Matter Interactions, University of Southern Denmark, Campusvej 55, DK-5230 Odense M, Denmark}

\date{\today}

\begin{abstract}

We investigate how the near-field enhancement of self-similar nanolenses,
made of three plasmonic nanospheres with decreasing sizes and separations,
is affected by quantum corrections in the optical response of the metal,
and by fabrication imperfections related to the positioning of the spheres
in the nanolens. In particular, we integrate mesoscopic phenomena, such as
electron spill-in and -out and surface-enabled Landau damping, via the
surface-response formalism, focusing particularly on the role of spill-out
in alkali metals. In addition, we take realistic imperfections in the
nanofabrication process into account, through numerical averaging of both
far- and near-field spectra for large collections of nanolenses. Statistical
analysis of hundreds of trimers shows that inevitable deviations from the
perfectly aligned chain only slightly, if at all, impair the field
enhancement, as long as the average misplacement is kept within 1 nm from
the ideal situation. Wishing to explore whether such imperfections can be
harvested for practical applications, we probe the potential for triggering
chiral response. Our results verify that imperfect nanolenses can display
considerable 
light-induced optical activity and
circular dichroism, while providing a means to manipulate the optical chirality
density. 
This highlights how promising the nanolensing effect is for chiral
light--matter interactions.
Nonetheless, we emphasize that quantification of chiral light--matter
interactions can be largely affected by mesoscopic phenomena, which cannot
be ignored when near-field quantities like optical chirality density are
investigated.

\end{abstract}

\maketitle

\section{Introduction}

Enhancing electromagnetic (EM) near-fields is tremendously important
for nanophotonic technologies such as increasing the sensitivity of
camera sensors~\cite{ho_etal_sciadv8}, improving the resolution of imaging
systems~\cite{koenderink_nanoph11} and enhancing the performance of
solar cells~\cite{burgt_acsphot7}. One of the most straightforward,
yet efficient methods to produce strong, highly-confined near fields,
is through plasmonic cavities, such as those formed by
nanosphere dimers~\cite{romero_optexp14,brown_acsnano4} and bowtie 
nanoantennas~\cite{huigao_nanolett12,santhosh_natcom7,carlson_prb102}.
Similarly to traditional Fabry--P{\'e}rot cavities, plasmonic fields
are trapped in the nanocavity created between nanoparticles (NPs),
thus magnifying the optical density of states in so-called EM 
hotspots~\cite{nielsen_optexp12,knebl_prb93,hardy_nscale16}. The
resulting field enhancement is undoubtedly valuable for surface-enhanced
Raman scattering (SERS)~\cite{xu_pre62,taley_nanolet5,langer_acsnano14},
fluorescence imaging~\cite{hoang_natcom6}, and plasmonic
photocatalysis~\cite{halas_natcom8}. By introducing additional NPs,
trimers and longer chains~\cite{barrow_nl11,tserkezis_ppsc31} are
produced, dominated by more complex hybrid plasmonic modes, and
providing more hotspots for field confinement.
 
Pursuing ever stronger near-field enhancement and localization, a
nanochain of three spheres with progressively decreasing radii and
separations was proposed several years ago~\cite{li_prl91}. Specifically,
within the quasistatic limit, a correlation between the radii of the
spheres and the interparticle gaps was suggested, to maximize the field
in the small gap as a consequence of the so-called plasmonic lensing
effect~\cite{bidault_jacs130,almpanis_optlet22,hoppener_prl109,joseph_acsaop2}.
Nevertheless, in reality, the quasistatic approximation is known to fail
at the mesoscopic scale, and quantum phenomena such as the spill-out or -in
of the induced electron density, surface-enabled Landau damping, and nonlocal
dynamical screening become relevant, calling for a departure from the
local-response approximation
(LRA)~\cite{zhu_natcom7,mortensen_nanoph10a,stamatopoulou_omex12}. 
The situation can become more unclear in collections of interacting
NPs, e.g., dimers and trimers, because of the hybridizations between
localized surface plasmons (LSPs) sustained by each NP, and how these
are affected by mesoscopic physics. To complicate things further,
experimental arrangements of NPs would deviate from the ideal system
of collinear NPs, possibly on the nanometer order. For instance,
electron beam lithography may be one of the most successful methods
for making ordered nanostructures~\cite{chen_micreng135}, but there
is still a degree of positioning uncertainty; the same can hold for
colloidal chemistry~\cite{boles_acscr116}, while DNA origami has
emerged as the most promising template for ordered plasmonic
systems~\cite{acuna_sci338}.

So far, nanolens trimers have been studied mainly with field confinement
in mind~\cite{li_prl91,bidault_jacs130,almpanis_optlet22,pellegrini_jpcc120, lloyd_acsnano11,joseph_acsaop2}
---possibly for applications in fluorescence
enhancement~\cite{tserkezis_prb96}--- for perfectly arranged chains
excited by linearly polarized light along the chain axis. Nevertheless,
the moment the NPs are displaced, and particularly in the direction
normal to both the chain axis and the incident wavevector, there is
a considerable probability for the system to acquire a chiral character.
For example, extensive investigation has been performed in the past for
trimeric nanospheres~\cite{chuntonov_mrsbull38},
nanorods~\cite{shen_chemcom51, chen_acsaplmat10, wang_biosen84} and
nanosphere pyramids~\cite{mastroianni_jacs132}, where the chiral response
of NPs still lying in the same plane is due to the circularly-polarized
excitation and the interaction between LSPs which rotate the polarization
of the scattered light, rather than due to the geometric chirality of the system.
Optical chirality has attracted attention in the fields of
nanophotonics~\cite{hentschel_sciadv3,stamatopoulou_nscale14,lininger_advmat35, baranov_acsphot10} and quantum optics~\cite{lodahl_nat541,barik_prb101}, as
it embodies a light--matter interaction that has still many surprises to
reveal. Generally, the interaction of chiral light with either achiral or
chiral nanostructures offers great possibilities for biophotonics and
quantum technologies. For instance, in drug development~\cite{brooks_ctmc11}
it is important for human safety to ensure the formation of the correct
enantiomeric chiral compound. When it comes to quantum technology,
manipulation of chiroptical properties can lead to fascinating applications
in quantum information processes~\cite{mahmoodian_prl117} and chiral
sensing~\cite{mohammadi_acsphot5}; chirality in molecular spins can also
be beneficial for quantum computing~\cite{chiesa_advmat35}. At the same
time, on a more fundamental level, open questions still exist with respect
to concepts such as the physical meaning of EM chirality
density~\cite{philbin_pra87,droulias_prl122,mun_lsa9,letsios_lmp113,trigo_prr6}. 

Here, we focus on a trimeric nanolens made of sodium (Na) nanospheres,
taking into account nonlocal and quantum phenomena in the NPs through the
surface-response formalism (SRF)~\cite{feibelman_pss12,yan_prl115,christensen_prl118,yang_nat576,mortensen_nanoph10b,chen2024},
with the intention to explore, in a self-consistent manner, the influence of
both electron spill-out and Landau damping. We perform a large series of
simulations within both LRA and SRF, for small variations of the NP
coordinates, to address the question about the importance of fabrication
imperfections and how robust the system might remain.
With such positioning deviations (asymmetric arrangement), we investigate
whether the emerging extrinsic chirality~\cite{lu_nanoscale6} leads to
chiroptical response, by examining perturbed nanolenses under left/right-
circular polarization (L/RCP).
We concentrate on the impact of mesoscopic quantum effects on the field of
chiral light--matter interactions~\cite{tserkezis_acsphot5}, and find that
imperfections can indeed be exploited to produce considerable circular
dichroism (CD)~\cite{yannopapas_optlett34,fan_jpcc115, yao_rsc10} and
alter the optical chirality
density~\cite{tang_prl104,corbaton_prx6,pham_pra98, poulikakos_symmetry11}.
In all cases, we underline that these quantities are severely influenced
by nonlocality and quantum effects.

\section{Methodology}\label{Sec:method}

\begin{figure}[h]
\centering
\includegraphics[width=0.57\columnwidth]{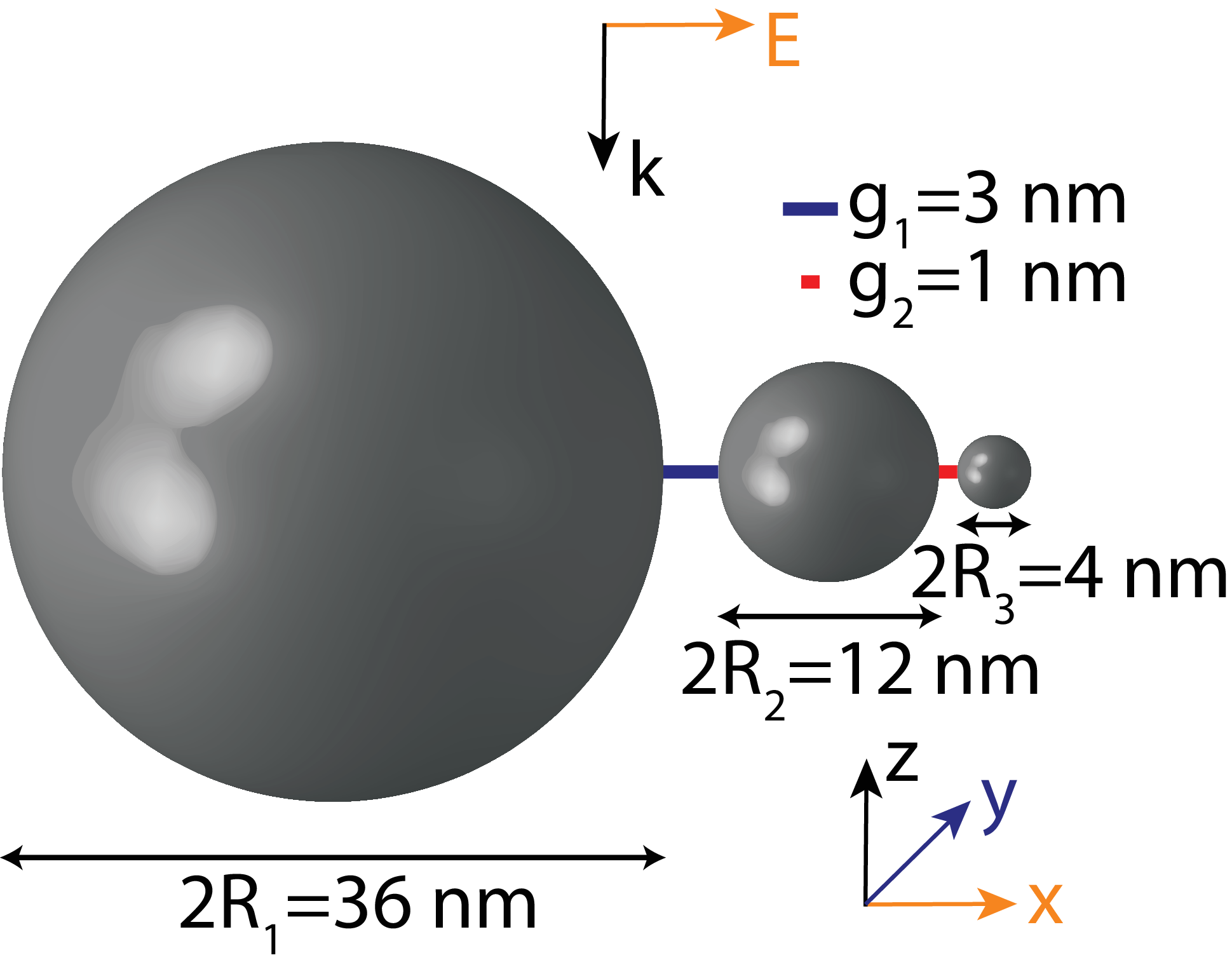}
\caption{
Sketch of a perfectly aligned plasmonic nanolens, and the NP and
gap dimensions considered throughout the paper. In the
first part of the paper, the lens is illuminated by a plane wave
linearly polarized along the chain axis ($x$ axis), as indicated
in the sketch.}
\label{Na_chain}
\end{figure}

The ideal, non-perturbed nanolens considered throughout the paper
is depicted in Fig.~\ref{Na_chain}: three nanospheres of decreasing
radii $R_{1} - R_{3}$ are placed at decreasing distances from each
other with interparticle gaps of lengths $g_{1}$ and $g_{2}$,
positioned along a straight line. To ensure the most efficient
excitation of chain modes, the polarization of the incident light
is parallel to the chain, i.e., along the $x$ axis, and the EM wave
propagates along the $z$ axis. EM hotspots will form in gaps
$g_{1}$ and $g_{2}$, with the field becoming more localized as the
spheres decrease in size. In what follows, we want to focus on how 
mesoscopic corrections affect this field enhancement.
Generally, in good, simple Drude metals like Na, electron spill-out
dominates. On the other hand, in noble metals like gold, d-electron
screening pushes the induced charge inwards. The role of spill-in and
screening is known~\cite{tserkezis_prb96}; here we want to focus on
spill-out and explore how the extension of the NP boundaries towards
the hot spot affects the performance of the nanolens.
To this end, we choose the plasmonic material to be Na, even though
such NP chains are challenging to achieve and stabilize in experiments.
The metal can be described by the local Drude permittivity,
\begin{equation}\label{Eq:Drude}
\varepsilon(\omega) = 
\varepsilon_\infty - 
\frac{\omega_{\mathrm{p}}^{2}}
{\omega (\omega + \mathrm{i} \gamma)}
,
\end{equation}
where $\omega$ is the angular frequency of the incident light. 
Throughout the paper, the plasma energy in eq~\ref{Eq:Drude} is taken
equal to $\hbar\omega_{\mathrm{p}} =5.89$\,eV,
and the damping rate $\gamma$ corresponds to energy $\hbar \gamma =
0.1$\,eV. For alkali metals, with a single free electron in the outer
orbital, the contribution of bound electrons and ions is simply 
$\varepsilon_{\infty} = 1$. Regarding dimensions, we choose the
largest sphere to have a radius of $R_{1} = 18$\,nm, the middle
sphere $R_{2} = 6$\,nm, and the smallest one, located at the end
of the chain, $R_{3} = 2$\,nm. The gap between the large and middle
spheres is $g_{1} =3$\,nm, whereas the gap between the middle and
small spheres is $g_{2} = 1$\,nm. Thus, the reduction in radii and
gaps follows a common geometrical progression with a common ratio
of $3$ which, according to our simulations, produces the optimal
field enhancement for this $R_{1}$ as the starting point. These
choices were motivated by the requirement to restrict $g_{2}$ outside
the tunneling regime~\cite{esteban_natcom3,yan_prl115}, while the
NP sizes should ensure that all individual LSPs are close to the
quasistatic limit and resonate at almost the same energies.

It is known that LRA is inadequate for the description of plasmonic
NPs of sizes comparable to intrinsic electron motion scales, like the
Fermi wavelength, due to limitations of the local scalar dielectric
function of the bulk material~\cite{teperik_prl110,raza_natcom6}.
As a first step, nonlocality can be addressed through the hydrodynamic
Drude model (HDM)~\cite{halevi_prb51,ruppin_prl31,abajo_jpcc112,Zheng_ats5}
and its extensions~\cite{mortensen_natcom5,ciraci_prb93,toscano_natcom6},
where the metal is a free-electron plasma described by two distinct
permittivity components, transverse ($\varepsilon_\mathrm{t}$) and
longitudinal ($\varepsilon_\mathrm{l}$),
\begin{align}
\varepsilon_{\mathrm{t}} (\omega) = 
\varepsilon_{\infty} - 
\frac{\omega_{\mathrm{p}}^{2}}
{\omega (\omega + \mathrm{i} \gamma)}
,\quad \,
\notag\\
\varepsilon_{\mathrm{l}} (\omega, \mathbf{k}) = 
\varepsilon_{\infty} - 
\frac{\omega_{\mathrm{p}}^{2}}
{\omega (\omega + \mathrm{i} \gamma) - 
\beta^{2} k_{\mathrm{l}}^{2}}
.
\label{dielectric}
\end{align}
On the right-hand side of eq~\ref{dielectric},
$\mathbf{k}$ is the full wavevector, $\beta$ is the hydrodynamic
parameter and $k_{\mathrm{l}}$ is the longitudinal wave number. Based
on the Thomas--Fermi theory, we adopt $\beta^{2} = \frac{3}{5}
v_{\mathrm{F}}^{2}$~\cite{halevi_prb51,wegner_prb107}, where
$v_{\mathrm{F}} = 1.39 \times 10^{6}\,\mathrm{m/s}$ denotes the Fermi
velocity of Na. Nonlocal phenomena are incorporated into HDM through
the dispersion relation $\varepsilon_{\mathrm{l}} = 0$, which implies
the existence of longitudinal waves in the bulk. This requires an
additional boundary condition, which is usually chosen as the
vanishing of the normal component of the induced current (hard-wall
boundary condition)~\cite{raza_jpcm27}.

Even with these modifications, the theoretical predictions of HDM
diverge from time-dependent density-functional theory (TDDFT)
predictions~\cite{teperik_oex21}, and the model is only used here
to highlight this inadequacy. This issue stems exactly from assuming
hard-wall boundary conditions. Actually, the induced charge has a
density distribution extending inward and outward from the NP surfaces,
commonly referred to as spill-in and spill-out, respectively, while
also exhibiting Friedel oscillations near the edge of the metal.
This specific extension of the charge density is significant in the
case of alkali metals, which are characterized by low work function.
To address this situation, we apply the more reliable method of
SRF~\cite{feibelman_pss12,christensen_prl118,goncalves_natcom11} which,
being based on surface-response functions, has the additional benefit
of accounting for surface-enabled Landau
damping~\cite{christensen_prl118,mortensen_nanoph10a}. The incorporation
of Feibelman $d$-parameters, 
\begin{align}\label{Eq:Feibelman}
d_{\perp} (\omega) = 
\frac{\int_{-\infty}^{\infty} z \rho(z) \, dz}
{\int_{-\infty}^{\infty} \rho(z) \, dz}, 
\quad
d_{\parallel} (\omega) = 
\frac{\int_{-\infty}^{\infty} z \, 
\partial_{z} J_x(z) \, dz}
{\int_{-\infty}^{\infty} \partial_{z} 
J_{x} (z) \, dz}
\end{align}
which quantify the centroids of the induced charge density along the
direction normal to the metal--environment interface ($z$),
$\rho (\mathbf{r}) \equiv \rho(z) \mathrm{e}^{\mathrm{i} qx}$, and
the corresponding tangential induced current density $\mathbf{J(r)} 
\equiv \mathbf{J}(z) \mathrm{e}^{\mathrm{i}qx}$, introduces
discontinuities in the parallel ($x$) components of the electric
and magnetic field (here, $q$ is the wavevector component parallel
to the interface). The real parts of the $d$-parameters, which are
complex numbers with units of length, indicate the location of the
charge or current centroid. Their imaginary components, meanwhile,
reflect the effect of surface-induced Landau damping.

A revised set of boundary conditions, to capture this blurring 
of the surface, is necessary. Using macroscopic Maxwell equations,
it has been shown~\cite{goncalves_natcom11} that the boundary conditions
\begin{align}
\mathbf{E}_{\mathrm{out}, \parallel} - 
\mathbf{E}_{\mathrm{in}, \parallel} &= 
-{d_\perp} \mathbf{\nabla}_{\parallel} \times 
(\mathrm{E_{\mathrm{out}, \perp}} - 
\mathrm{E_{\mathrm{in},\perp}})
, \label{Eq:dBoundary1} \\
\mathbf{H}_{\mathrm{out}, \parallel} - 
\mathbf{H}_{\mathrm{in}, \parallel} &= 
\mathrm{i} \omega d_{\parallel} 
(\mathbf{D}_{\mathrm{out}, \parallel} - 
\mathbf{D}_{\mathrm{in}, \parallel}) \times 
\mathbf{\hat{n}}
, \label{Eq:dBoundary2}
\end{align}
reproduce the discontinuities due to surface polarization and
current, which are proportional to the $d$-parameters
of eq~\ref{Eq:Feibelman}.

The interaction of the incident plane wave with our system is studied
via the so-called $\mathbf{T}$-matrix method~\cite{waterman_ieee53}.
The main equation behind the method is,
\begin{equation}\label{Tmatrix0}
\mathbf{f}_{i} = 
\mathbf{T}_{i} \cdot \left[ 
\mathbf{f}_{i}^{\mathrm{inc}} + 
\sum_{j = 1, j \not= i}^{j=3} 
\mathcal{T}^{13} (\mathbf{r}_{ji}) \cdot 
\mathbf{f}_{j} \right]
.
\end{equation}
In eq~\ref{Tmatrix0}, the subscripts $i$ and $j$ refer to the
$i^\mathrm{th}$ and $j^\mathrm{th}$ NP, with $i$ and $j$ spanning
from $1$ to $3$. $\mathbf{f}_{i}$ is a column vector
$\begin{pmatrix} \mathbf{a}_{i}, \mathbf{b}_{i} \end{pmatrix}^{T}$
where $T$ marks the matrix transpose, and $\mathbf{a}_{i}$ and
$\mathbf{b}_{i}$ are also row vectors containing expansion coefficients
$a_{nm}$ and $b_{nm}$ of the scattered field $\mathbf{E}_{s} (\mathbf{r})$,
re-radiated by the NP centered at $\mathbf{r}_{i}$,
\begin{equation}\label{Tmatrix1}
\mathbf{E}_{s} (\mathbf{r}) = 
\sum_{n,m} \mathbf{M}_{nm}^{(3)} (\mathbf{r} - \mathbf{r}_{i}) a_{mn} + 
\frac{1}{\mathrm{i} Z} 
\mathbf{N}_{nm}^{(3)} (\mathbf{r} - \mathbf{r}_{i}) b_{mn}
.
\end{equation}
On the left-hand side of eq~\ref{Tmatrix1}, $\mathbf{r}$ is a field
point outside all the NPs. On the right-hand side of the equation, the
summation is done with respect to two integer indices $n$ and $m$.
They are known as the angular momentum quantum number, and the
magnetic quantum number, with the former and the latter spanning
from $1$ to $\infty$ and from $-n$ to $+n$, respectively. Furthermore,
$\mathbf{M}_{nm} (\mathbf{r} - \mathbf{r}_{i})$ and
$\mathbf{N}_{nm} (\mathbf{r} - \mathbf{r}_{i})$ are the vector spherical
wave functions~\cite{Chew_EM}. The superscript $(3)$ denotes spherical
waves of radiating type; the spherical Hankel function must be used in
the wave functions. We underline that $\mathbf{r} - \mathbf{r}_{i}$
emphasizes that the spherical waves are expanded with respect to the
center of the $i^\mathrm{th}$ sphere, i.e., $\mathbf{r}_i$. Lastly,
$Z$ is the wave impedance of the material filling the space outside
all NPs.

Back to eq~\ref{Tmatrix0}, similar to $\mathbf{f}_i$,
$\mathbf{f}_i^{\text{inc}}$ is also a column vector
$\begin{pmatrix} \mathbf{a}_{i}^{\text{inc}},
\mathbf{b}_{i}^{\text{inc}} \end{pmatrix}^{T}$. The difference is that
the row vectors $\mathbf{a}_{i}^{\text{inc}}$ and
$\mathbf{b}_{i}^{\text{inc}}$ include the expansion coefficients
$a_{nm}^{\text{inc}}$ and $b_{nm}^{\text{inc}}$ of the incident plane
wave $\mathbf{E}_{\text{inc}} (\mathbf{r})$,
\begin{equation}\label{Tmatrix2}
\mathbf{E}_{\mathrm{inc}} (\mathbf{r}) = 
\sum_{n,m} \mathbf{M}_{nm}^{(1)} (\mathbf{r} - \mathbf{r}_{i})
a_{mn}^{\mathrm{inc}} + 
\frac{1}{\mathrm{i} Z} 
\mathbf{N}_{nm}^{(1)} (\mathbf{r} - \mathbf{r}_{i}) 
b_{mn}^{\mathrm{inc}}
.
\end{equation}
In eq~\ref{Tmatrix2}, the superscript $(1)$ indicates spherical wave functions of
standing type, and spherical Bessel functions are used. To complete
the discussions on eq~\ref{Tmatrix0}, on the one hand, we note that
$T_i$ is the so-called transition matrix of the $i^\mathrm{th}$
particle. The evaluation of this method has been thoroughly discussed
in previous works~\cite{mystilidis_cphysc284,yan_ieeetmtt72}. It should
be underscored that $T_{i}$ is the \emph{only} element that is affected
by the material models. On the other hand,
$\mathcal{T}^{13} (\mathbf{r}_{ji})$, where $\mathbf{r}_{ji}$ is a
displacement vector pointing from the center of the $i^\mathrm{th}$ NP
to the one of the $j^\mathrm{th}$ and the superscript marks a transformation
from the radiating spherical waves to the standing spherical waves, is the
translation matrix~\cite{Chew_EM} and is to reconcile the difference in
the expansion centers of the spherical wave functions. Here, we remember
that $\mathbf{f}_{i}$ and $\mathbf{f}_{j}$ contain the expansion
coefficients of the wave functions centered at $\mathbf{r}_{i}$, and
centered at $\mathbf{r}_{j}$. The detailed expressions for the
translation matrix can be found in Appendix~D of Chew's
textbook~\cite{Chew_EM} and will not be reproduced here. Finally,
after listing eq~\ref{Tmatrix0} for each NP, we reach a system of
matrix equations, and by solving the equations, we obtain the expansion
coefficients $\mathbf{f}_{i}$, which allows to evaluate the scattered
field $\mathbf{E}_{\mathrm{s}}$ at any point $\mathbf{r}$ in space, and
thus all the associated physical quantities, e.g., the scattering and
absorption cross sections. Different mesoscopic models can be
straightforwardly incorporated in the $\mathbf{T}$-matrix method,
simply by introducing he corresponding single-particle Mie coefficient,
either for HDM~\cite{ruppin_prl31} via the inclusion of longitudinal
waves, or for SRF~\cite{eriksen_nanoph11} via the modified boundary
conditions of eqs~\ref{Eq:dBoundary1}-\ref{Eq:dBoundary2}.

\section{Results and discussion}

\subsection{Nanolens under different EM models}

\begin{figure*}[!ht]
\centering
\includegraphics[width=0.77\linewidth]{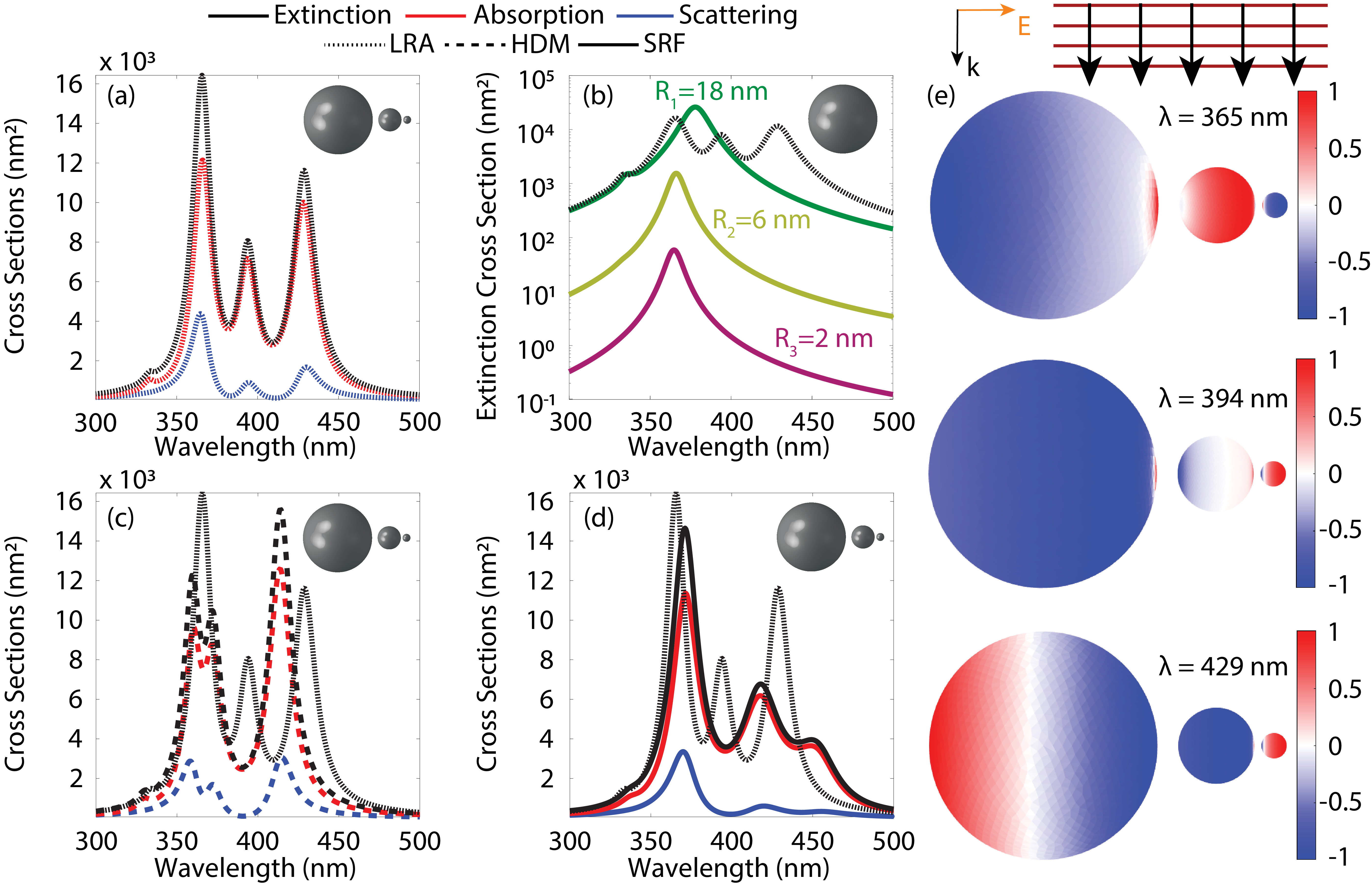}
\caption{
(a) Extinction (black color), scattering (blue color) and absorption
(red color) cross section for a Na nanolens, in the case of linearly
polarized light, within the context of LRA (dotted lines).
(b) Extinction cross section (logarithmic scale) of each individual
sphere (solid lines), compared to the extinction of the full trimer
(dotted line).
(c) Comparison of LRA (extinction only, dotted line) with the cross
sections calculated with HDM (dashed lines, same color code as in a).
(d) Same as (c), now with SRF (solid lines) as the mesoscopic model.
(e) Sketch of the surface charge distribution at the wavelengths of
the three chain resonances in (a), i.e, $\lambda=365$\,nm,
$\lambda=394$\,nm, and $\lambda=429$\,nm.
}
\label{cross_section}
\end{figure*}

Starting with LRA, the optical response of the ideal linear nanolens
is dominated by three main plasmonic resonances shown in 
Fig.~\ref{cross_section}a, where absorption governs the extinction
cross section, as expected for such small NPs. The extinction cross
sections of each sphere individually are compared in
Fig.~\ref{cross_section}b. As a consequence of retardation, the
large sphere exhibits a resonance at $378$\,nm, at a longer wavelength
compared to the medium-sized sphere at $366$\,nm, and the small sphere
at about $365$\,nm, which follow the quasistatic approximation and
resonate at $\omega_{\mathrm{p}}/\sqrt{3}$. When all three spheres
are arranged in the fully-formed nanolens, LSPs of the individual
spheres interact to form three new bonding LSPs. These chain modes
depend on how the individual polarizations of the spheres are coupled.
Specifically, we portray in Fig.~\ref{cross_section}e the charge
distributions on the NP surfaces at wavelengths $365$\,nm, $392$\,nm,
and $428$\,nm, respectively, in order to visualize the physical
mechanisms, i.e., bonding and anti-bonding modes that cause the
three resonances; here, the minimum and maximum charges are
normalized to $\pm 1$.

As a first departure from LRA, we turn our attention to nonlocal
phenomena as captured in Fig.~\ref{cross_section}c by HDM (dashed
lines). Longitudinal field components signify that induced charges
extend spatially, rather than being solely confined to the
metal--dielectric interface. This shift of the centroid of the
induced charge inwards means that the total restoring force in
plasmonic oscillations is increased, causing a blueshift of LSPs
compared to LRA predictions. The resonances maintain similar
magnitudes, since no additional damping mechanism has been added,
and the shifts are typically small, yielding only slight alterations
in the metal's dielectric. The chain modes are modified primarily
due to the smallest sphere, which is strongly affected by nonlocality. 

\begin{figure}[h]
\centering
\includegraphics[width=\columnwidth]{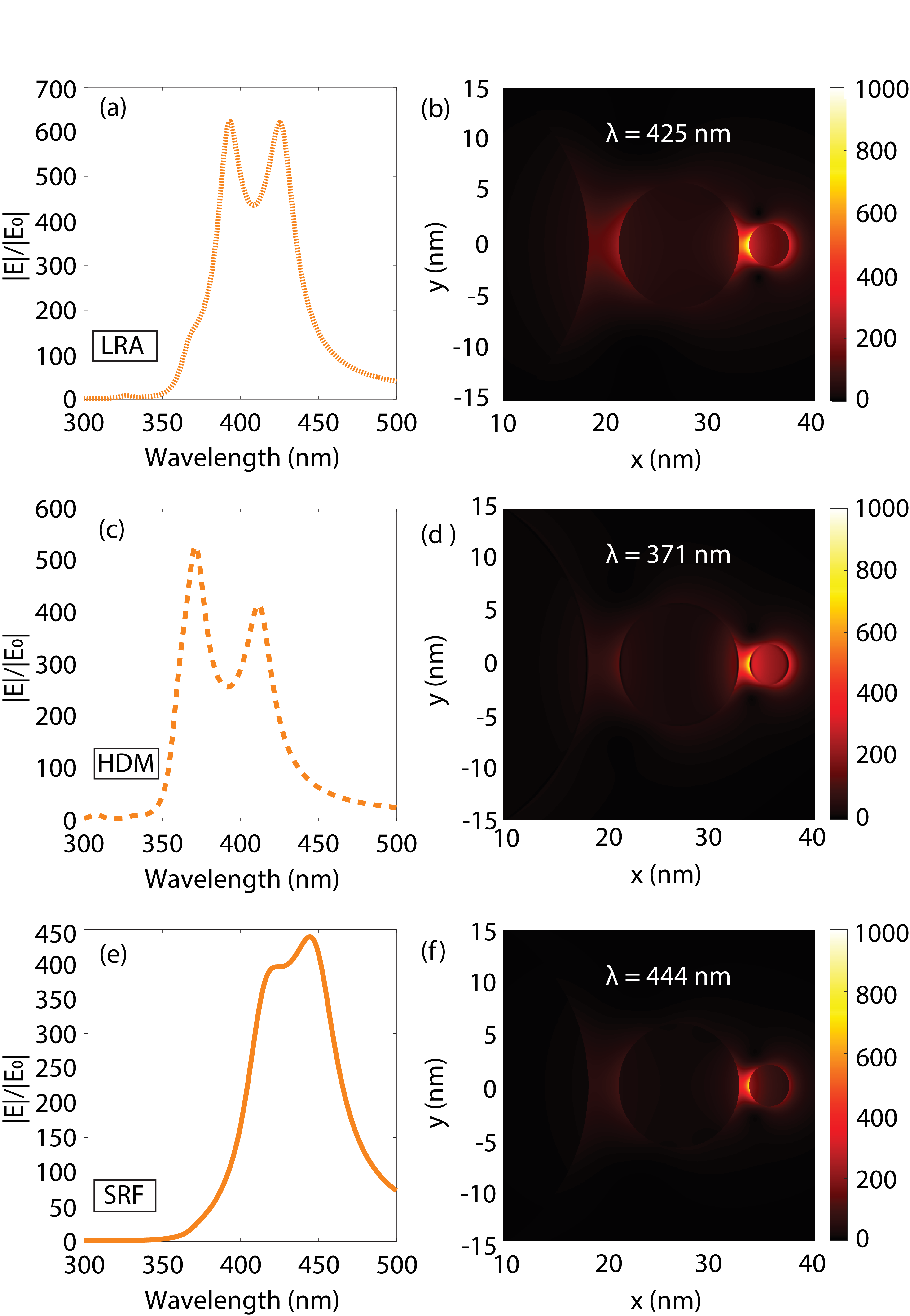}
\caption{
(a, c, e) Electric-field-enhancement spectra at the hotspot of $g_{2}$,
and (b, d, f) contour plots of the electric field at wavelengths
$\lambda = 425$\,nm, $\lambda = 371$\,nm and $\lambda = 444$\,nm,
respectively. Panels a and b correspond to LRA, c and d to HDM, and
e and f to SRF.
}
\label{electric_field}
\end{figure}

Although these results highlight the necessity to depart from LRA,
both \emph{self-consistent} HDM and TDDFT calculations, have indicated
that one should anticipate redshifts, instead of blueshifts of the
modes, owing to electron spill-out and the relaxation of the hard-wall
boundary condition~\cite{teperik_prl110,toscano_natcom6,ciraci_prb93}.
To efficiently capture spill-out, we now turn to SRF, and in particular
to the Feibelman parameters associated with the sodium--air 
interface~\cite{yan_prl115}. The obtained cross sections are illustrated
in Fig.~\ref{cross_section}d (solid lines). We observe that the chain
modes undergo a considerable inhomogeneous spectral broadening, which
is related to Landau damping~\cite{christensen_prl118,Uskov_JPCL2022},
and the resonances are redshifted, aligning with the predictions
of TDDFT calculations for single NPs and dimers~\cite{babaze_oex30}.
These results are in good agreement with a previous investigation
of the same system~\cite{fitzgerald_acsphot5}, where nonlocality
and electron spill-out were introduced through the local analogue
model of Luo et al.~\cite{luo_prl111}. It is important to note that,
in the wavelength range of $400-460$\,nm, we can detect a strong
damping of the chain modes owing to the increased surface-to-volume
ratio for the smaller NPs, and the dominance of surface
effects~\cite{eriksen_nanoph11}.

The main interest in plasmonic nanolenses is the behavior of the near
field in the smallest nanogap. In Fig.~\ref{electric_field}a,c,e
we depict the enhancement of the electric field
($|\mathrm{E}|/|\mathrm{E_0}|$) at the $g_{2}$ plasmonic hotspot for
LRA, HDM, and SRF respectively, where $\mathrm{E_0}$ is the incident
electric field, while Fig.~\ref{electric_field}b,d,f present near-field
contours in the nanocavities, at the wavelength of maximum enhancement.
Substantial enhancement of the field is achieved at both chain modes
within all three models, reaching a maximum around $620$ within LRA,
$530$ for HDM, and $440$ for SRF, as is illustrated in
Fig.~\ref{electric_field}a, Fig.~\ref{electric_field}c and
\ref{electric_field}e, respectively. The contours of
Fig.~\ref{electric_field}b,d,f present the enhancement of the
electric field at wavelengths $425$\,nm (Fig.~\ref{electric_field}b),
$371$\,nm (Fig.~\ref{electric_field}d), and $444$\,nm
(Fig.~\ref{electric_field}f), in the plane defined by the chain
axis ($x$ axis) and the wavevector ($z$ axis). It is evident that,
apart from the significant field enhancement at the wavelength region
of the chain modes, considerable lensing and focusing takes place
in the vicinity of the smallest NPs. These plots show that the
enhancement of the plasmonic fields remains considerable even when
HDM and SRF are adopted.

\subsection{Fabrication imperfections}

\begin{figure}[hb]
\centering
\includegraphics[width=\columnwidth]{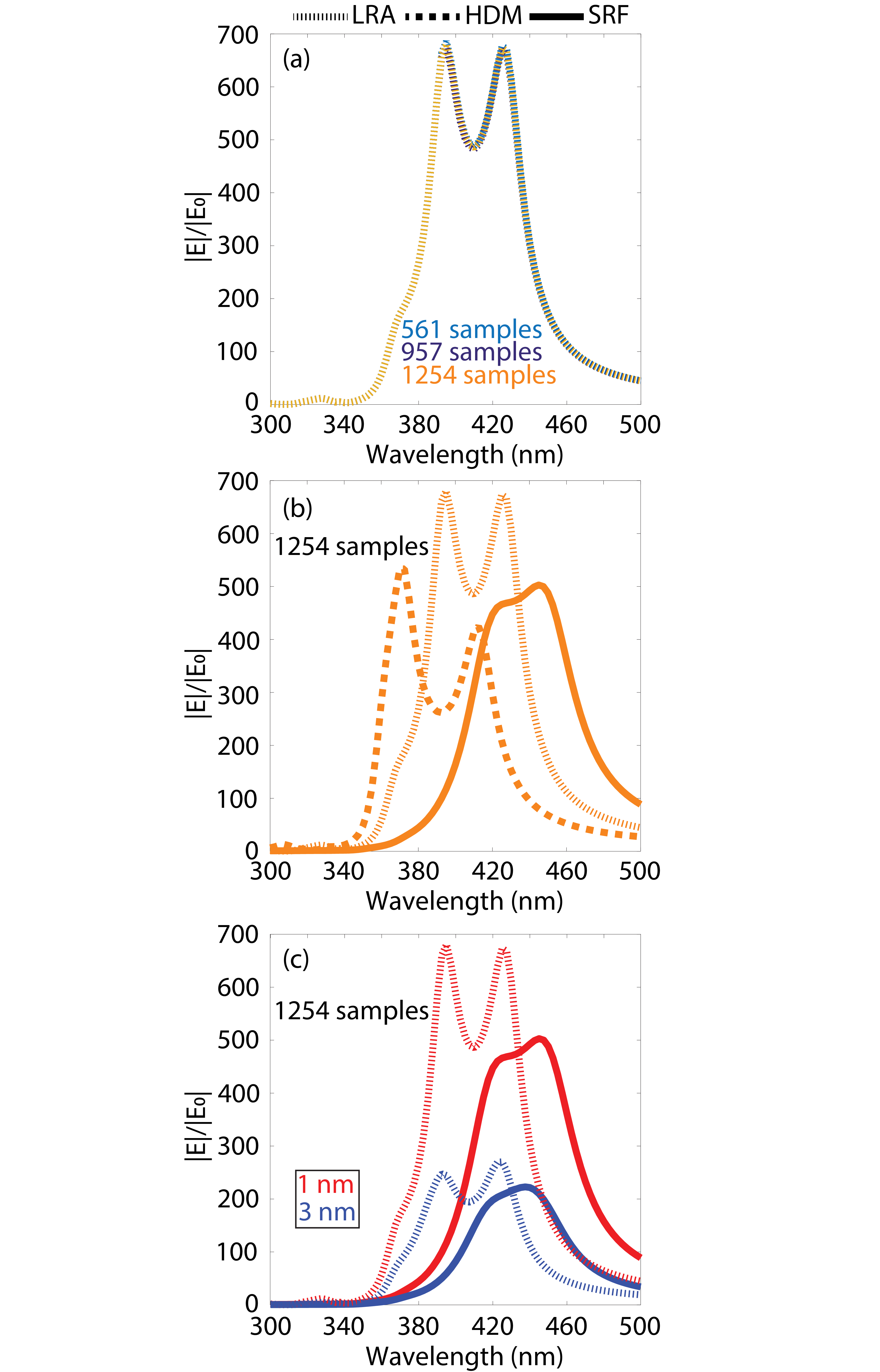}
\caption{
(a) Electric field enhancement for different numbers of nanolens
samples in the statistical collection, in the framework of LRA.
(b) Enhancement of the electric field under LRA (dotted lines),
HDM (dashed lines) and SRF (solid lines), and
(c) comparison of LRA and SRF for two different maximum displacements,
namely, $1$\,nm (red color) and $3$\,nm (blue color) in an ensemble of
$1254$ nanonlens realizations.
}\label{statistics}
\end{figure}

As already discussed, experimental preparation of the above system
is likely to have a series of imperfections. Both the radius of the
NPs and their separations are subject to small deviations. Similarly,
positioning the NP centers exactly on the same axis might be challenging.
It then makes sense to evaluate not the perfect sample, but the average
response of an ensemble. The effect of ensemble averaging for NP radii
and separations with different mesoscopic models has been studied in
the past~\cite{tserkezis_scirep6}.
Another practical imperfection concerns the surface roughness of the
individual NPs, which, in the particular case of plasmonic self-similars
chains, has been shown to not disrupt the nanolensing
effect~\cite{coluccio_sciadv1,kanehira_acsnano17}. Additionally, Wiener
et al.~\cite{wiener_nl12} have shown that nonlocality will smooth surface
imperfections in terms of field enhancement. Consequently, we expect
the response of perfectly aligned realistic NPs to closely align with
our calculations above.
We thus focus here on the more unknown role of NP positioning, maintaining 
the sizes and gaps fixed at the nominal values.

In Fig.~\ref{statistics} we show the statistical analysis for the
field enhancement at the $g_{2}$ hotspot, incorporating uncertainty
with respect to the position of the NPs. Inspired by our earlier work
on NP collections~\cite{tserkezis_scirep6}, we randomly modify the
central positions of the two smaller spheres. For our numerical
ensemble averaging, sufficiently large ensembles are generated to
ensure convergence of all ensemble-averaged quantities. Given the
semi-analytical nature of the problem, ensembles as large as 10$^5$
configurations, tested in our simulations for the field enhancement
at a single wavelength, remain treatable within only hours of
computation. To obtain the full picture, we present in
Fig.~\ref{statistics}a the LRA near-field spectrum, averaged for
three different numbers of different chain arrangements, sufficiently
large to ensure statistical convergence. The unchanged gap sizes lead
to fixed distances between NP centers, and allow us to vary the
relative directions among NPs. Since a direction can be parameterized
by the elevation and the azimuth angle, we can sample over a unit
sphere and the sampling is done by the so-called Kent 
distribution~\cite{mardia2009directional}, which is a spherical
equivalent of the two-dimensional (2D) normal distribution. In
the statistical analysis, we employ a standard deviation
corresponding to shifts by $1$\,nm away from the average positions.
The average enhancement of the electric field is depicted in
Figure~\ref{statistics}b for LRA (dotted lines), HDM (dashed lines),
and SRF (solid lines). Compared to Figure~\ref{electric_field}, the
averaged field enhancement remains close to the ideal linear case
within each model, signifying that deviations do not play a major
role. This could be anticipated by the fact that the lensing is
due to the smaller sphere interacting with a much larger one, and
the $1$\,nm deviation does not take it outside the ``shadow'' of
its neighbor. To check the limits of this performance, we implement
an extremely disadvantageous system where the displacement is as
large as $3$\,nm, keeping in mind that experiments typically offer
more accuracy. We can see in Fig.~\ref{statistics}c that the
outcome may not be as efficient as in the case of $1$\,nm, but
it is obvious that a very good field enhancement is still present
in both LRA and SRF. These findings imply that chain resonances
of imperfect plasmonic nanolenses
are robust and the field enhancement at the hotspot of the small
nanocavity is very strong even given realistic discrepancies.

\subsection{Chiral response of imperfect nanolenses}
In view of the aforementioned issues, the following question
arises: what happens when the NPs do not form a perfectly aligned
chain? Can the structure acquire a chiral character, and how strong
would that be compared to more conventional examples? 
While trimeric systems are not geometrically chiral, because they
lie in the same plane~\cite{hentschel_nanolet12} with respect to
which the mirror image will always be symmetric, a light-induced
chiral response, dependent on the incident wavevector of the
illumination, can arise through the interaction of LSPs in
arrangements with planar chirality~\cite{papakostas_prl90,gryb_nanolet23,gautier_acsphot9}.
So far, we inspected the optical response of linear or perturbed
trimeric nanolenses to linearly polarized plane-wave excitations,
but in the case of broken symmetry the handedness of circular polarization
plays a critical role. In what follows, we focus on optical chirality
density $C$ in the case of a
planar-chiral, imperfect arrangement of the three Na nanospheres (same radii
and gaps as before), with the intention of optimizing CD, and controlling
$C$ in the $g_{2}$ gap. CD constitutes one of the most frequently used
measures to estimate the chirality of 
a system;
it is defined as the differential absorption under the two different
circular polarizations, and is traditionally quantified through the
dissymmetry factor
\begin{equation}\label{Eq:CD}
\mathrm{CD} = 
\frac{\sigma_{\mathrm{abs}}^{\mathrm{LCP}} -
\sigma_{\mathrm{abs}}^{\mathrm{RCP}}}
{\sigma_{\mathrm{abs}}^{\mathrm{LCP}} +
\sigma_{\mathrm{abs}}^{\mathrm{RCP}}}
,
\end{equation}
where $\sigma_{\mathrm{abs}}^{\mathrm{L/RCP}}$ represents the absorption
cross section for L/RCP incidence. CD can originate from either the geometry,
the anisotropy of the material~\cite{rebholz_acsphot11}, 
or from extrinsic, light-induced chirality, when symmetry is broken for
the combined structure plus incident light system.
However, the CD spectrum alone cannot characterize the chirality of the
near field.

On the other hand, several years ago it was suggested by 
Lipkin~\cite{lipkin_jmathphys5} that the chirality of the EM fields is
correlated with an intrinsic symmetry that gives rise to a new conservation
law~\cite{ibragimov_aam105,cameron_newjphys14} for Maxwell's equations.
Lipkin introduced the following time-even pseudoscalar quantity
\begin{align}\label{C}
C = 
&\frac{1}{2} 
\left[ \mathbf{E} \cdot 
\left( \boldsymbol{\nabla} \times \mathbf{D}\right) + 
\mathbf{H} \cdot 
\left( \boldsymbol{\nabla} \times \mathbf{B} \right)\right] \notag \\
=
&\frac{\omega \varepsilon \mu}{2 c^{2}}
\mathrm{Im} \left[
\mathbf{E} (\mathbf{r}, \omega) \cdot
\mathbf{H}^{\ast}  (\mathbf{r}, \omega) \right]
,
\end{align}
where the first expression contains the time-dependent electric
($\mathbf{E}$), displacement ($\mathbf{D}$), magnetic ($\mathbf{H}$),
and induction ($\mathbf{B}$) fields, while the second expression gives
the time-averaged chirality density in a lossless medium described by
permittivity $\varepsilon$ and permeability $\mu$ (here, $c$ is the
speed of light in vacuum). Interest of the photonics community in this
quantity increased dramatically after the suggestion of Tang and 
Cohen~\cite{tang_prl104} that $C$, now called the optical
chirality (volume integral of eq~\ref{C}), is proportional to the
excitation rates of a molecule and can be used to quantify the
chirality of the near fields.
Additional measures have been proposed to find correlations between the
geometry of chiral objects (or anisotropic materials),
the interaction-induced effects,
and the degree of chirality and swirling of an EM field and, more
generally, what causes the asymmetry, but whether this is indeed a
fundamental quantity remains an open discussion~\cite{mun_lsa9}.
As we will show next, while CD is only present for chiral structures,
$C$ is neither given nor can be predicted based on geometry. Specifically,
the optical chirality density $C$ can be enhanced compared to vacuum
even for a straight chain of nanospheres; by changing the configurations
of the trimeric nanolens, we modify the scattered fields and can possibly
obtain the desired phase difference between the electric and magnetic
fields, $\Delta \phi \simeq \pi/2$, 
along with parallel, or nearly parallel/antiparallel fields and spatial
overlap .
Therefore, a proper arrangement of the NPs is necessary to allow the
chiral interaction of individual plasmonic modes within the successive
gaps.

\begin{figure*}[!ht]
\centering
\includegraphics[width=0.8\linewidth]{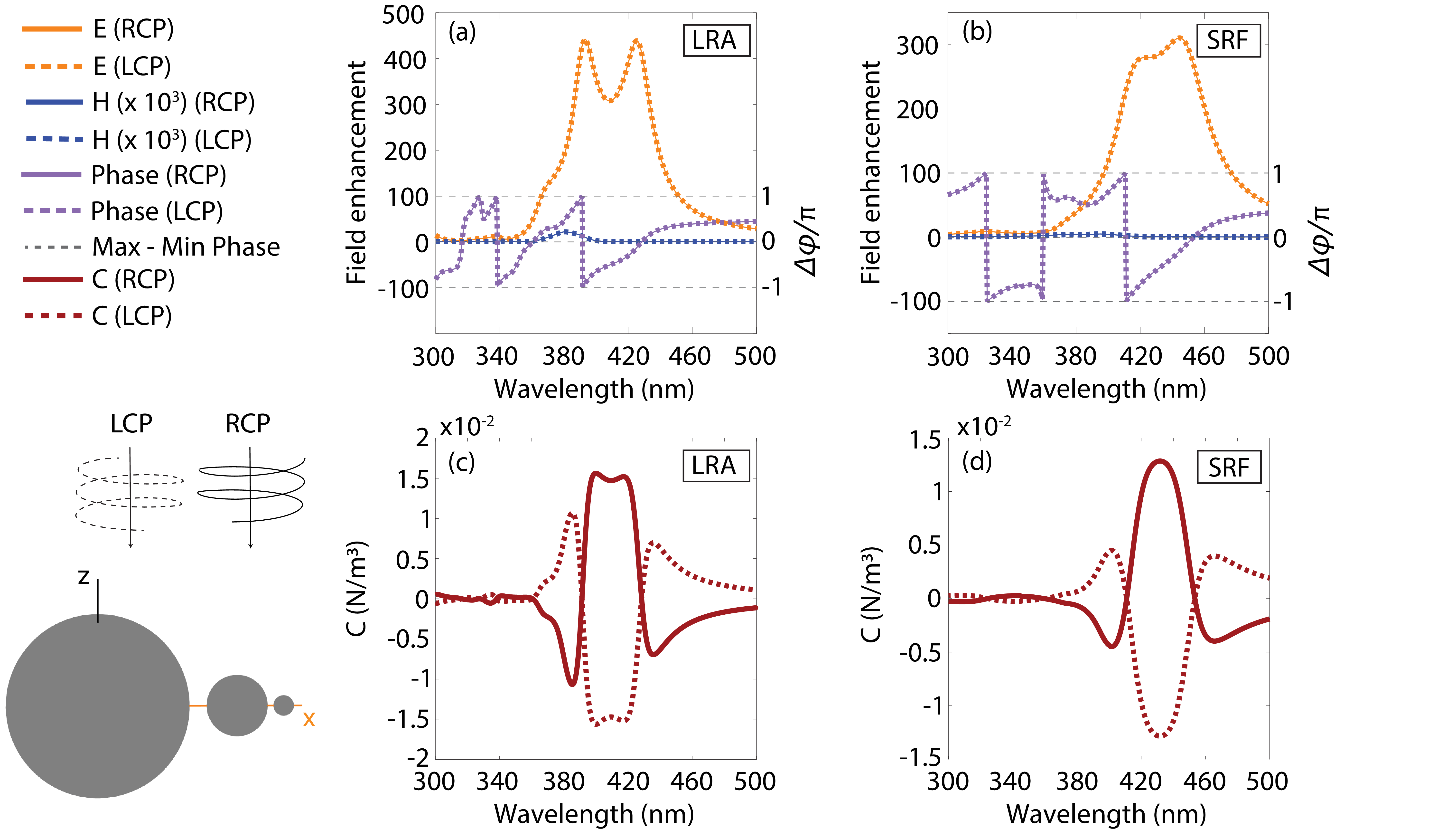}
\caption{
(a, b) Enhancement of electric field (orange color, left-hand axis),
magnetic field (blue color, left-hand axis), for a linear Na nanolens,
along with their respective phases (purple line, right-hand axis) at
the $g_{2}$ hotspot, in the case of RCP (solid lines) and LCP (dashed
lines), within the context of (a) LRA and (b) SRF. (c, d) Optical
chirality density $C$ in the middle of gap $g_{2}$, within (c) LRA and (d)
SRF, for R/LCP incident light.}
\label{figure4}
\end{figure*}

At first, we consider the scenario of a linear chain of Na nanospheres
(Fig.~\ref{figure4}) and compare LRA (Fig.~\ref{figure4}a,c) and SRF
(Fig.~\ref{figure4}b,d) to estimate the significance of the electron
density tail in chiral light--matter interactions. In Fig.~\ref{figure4}a,b
we illustrate the enhancement of the electric field $E$ (orange line) and
magnetic field $H$ (blue line), along with their respective phase difference
(purple line, right-hand axis) at the $g_{2}$ hotspot. In
Fig.~\ref{figure4}c,d we calculate $C$ (red line), exactly at the center
of $g_{2}$, for chiral light, both RCP (solid lines) and LCP (dashed lines).
As in the previous examples, the incident light propagates along the $z$ axis.
We can see that, despite not dealing with a chiral geometry, both EM fields
and $C$ increase. Obviously, due to the chain symmetry, whether it is RCP
or LCP light does not matter, and the enhancement of the fields (about
$450$ times) is identical, with the sign of $C$ being the only difference. 

\begin{figure*}[!ht]
\centering
\includegraphics[width=0.9\linewidth]{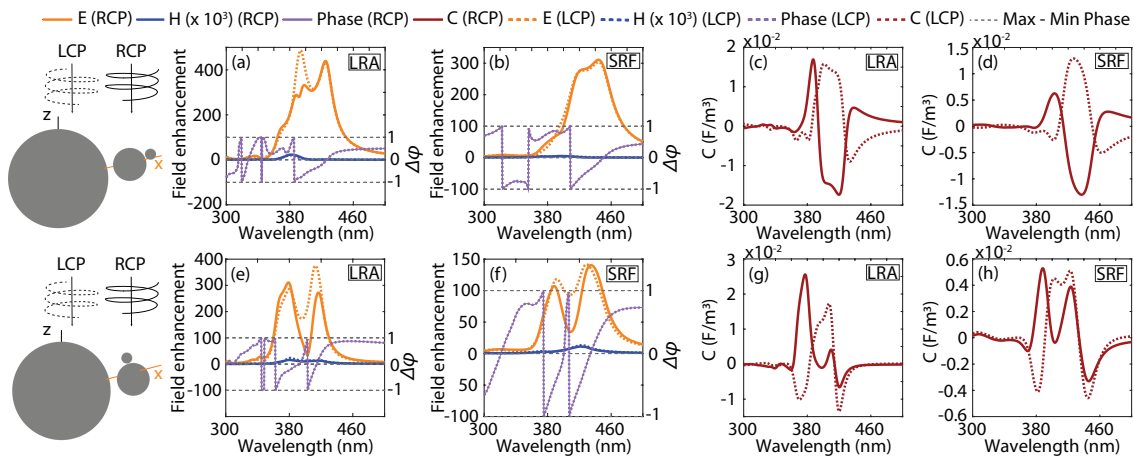}
\caption{
(a, b, c, d) Same as Fig.~\ref{figure4} (a, b, c, d) for Chiral
Chain A (depicted on the upper left-hand sketch) and (e, f, g, h)
for Chiral Chain B (depicted on the lower left-hand sketch). The
$x$ axis appears rotated in the sketch for illustrative reasons.
}
\label{figure5}
\end{figure*}

Subsequently, we displace the center of the second sphere by $1$\,nm
along the $y$ axis, and the center of the third sphere by $-1$\,nm
along the $y$ axis and $1$\,nm along the $z$ axis. Regarding this
specific arrangement, which we will refer to as Chiral Chain A,
in Fig.~\ref{figure5}a,b,c,d we evaluate the same physical
quantities as in Fig.~\ref{figure4} for the straight chain. In
Fig.~\ref{figure5}e,f,g,h we also consider larger NP displacements,
to examine what happens when we completely disengage from the symmetry
of the straight line. We call this arrangement Chiral Chain B, and it
is obtained by shifting the center of the second sphere by $4$\,nm
along the $y$ axis, and the center of the third sphere by $-4$\,nm
along the $y$ axis and $4$\,nm along the $z$ axis, respectively.
The displacements on the $x$ axis naturally are adjusted so that the
gap lengths $g_{1}$ and $g_{2}$ remain consistent with their initial
definitions in the beginning of the section.
We stress again that the names ``Chiral Chain`` reflect the
chiroptical response of the systems, and do not imply any intrinsic
geometrical chirality.

We notice that, whether we have an achiral 
system
or a chiral one, it appears that the maximum $C$ occurs when the
chain eigenmodes, where the intensity of the EM fields are at their
peak, coincide with a nearly-$\pi/2$ EM phase difference. Hence, $C$ is
determined by the phase difference between the electric and magnetic
fields. It can been seen that it is necessary for the nanostructure
to support both electric and magnetic field enhancement simultaneously
for $C$ to be non-negligible. On the other hand, in the comparison of
Fig.~\ref{figure4} with Fig.~\ref{figure5}a,b,c,d, we realize
that, even for small deviations of the NPs (chiral system), the
chiral character of the trimer emerges and the spectral profile of
the fields and the optical chirality density is completely altered, dictated
by the helicity of light.

Using circularly polarized light, it is feasible to control and
increase the optical chirality density in the vicinity of $g_{2}$,
reaching values around $C \sim -1.8 \times 10^{-2}$ at about $420$\,nm
(LCP) for Chiral Chain A, and $2.6 \times 10^{-2}$ at about $375$\,nm
(RCP) for Chiral Chain B, as shown in Fig.~\ref{figure5}c and
\ref{figure5}g, respectively, while the maximum value for the
linear chain is almost $1.6 \times 10^{-2}$ ($-1.6 \times 10^{-2}$) at
$400$\,nm for RCP (LCP), as depicted in Fig.~\ref{figure4}c.
Nevertheless, when quantum corrections are included within SRF,
we observe notable modifications in Fig.~\ref{figure4}b,d,
Fig.~\ref{figure5}b,d and Fig.~\ref{figure5}f,h. Decrease
of the optical chirality density is anticipated, because of the
weaker field enhancement, compared to LRA, owing to the damping
of the chain modes. In particular, $C$ assumes values as high as
$-1.3 \times 10^{-2}$ at $437$\,nm (LCP) for Chiral Chain A, and
slightly above $0.5 \times 10^{-2}$ at $388$\,nm (RCP) for Chiral
Chain B. It should be noted that applying SRF in a 
(planar-)chiral geometry, even if the optimal phase 
($-\pi/2$ or $+\pi/2$)
coincides with the eigenfrequencies, $C$ does not exceed the
amplifications obtained in the straight arrangement within the LRA
framework, because of the remarkable weakening of the fields.
Furthermore, by comparing Fig.~\ref{figure5}a,b and
Fig.~\ref{figure5}e,f, in light of LRA or SRF, we conclude
that Chiral Chain A enhances the near-field spectrum more than
Chiral Chain B. However, employing LRA, Chiral Chain B causes a
greater increase of $C$ at the $g_{2}$ hotspot whereas under the
influence of mesoscopic effects, the opposite is true, as $C$
decreases substantially due to the damping mechanisms, and at the
same time, the shifting in eigenfrequencies does not align with
the ideal phases of the fields.

\begin{figure}[h]
\centering
\includegraphics[width=\columnwidth]{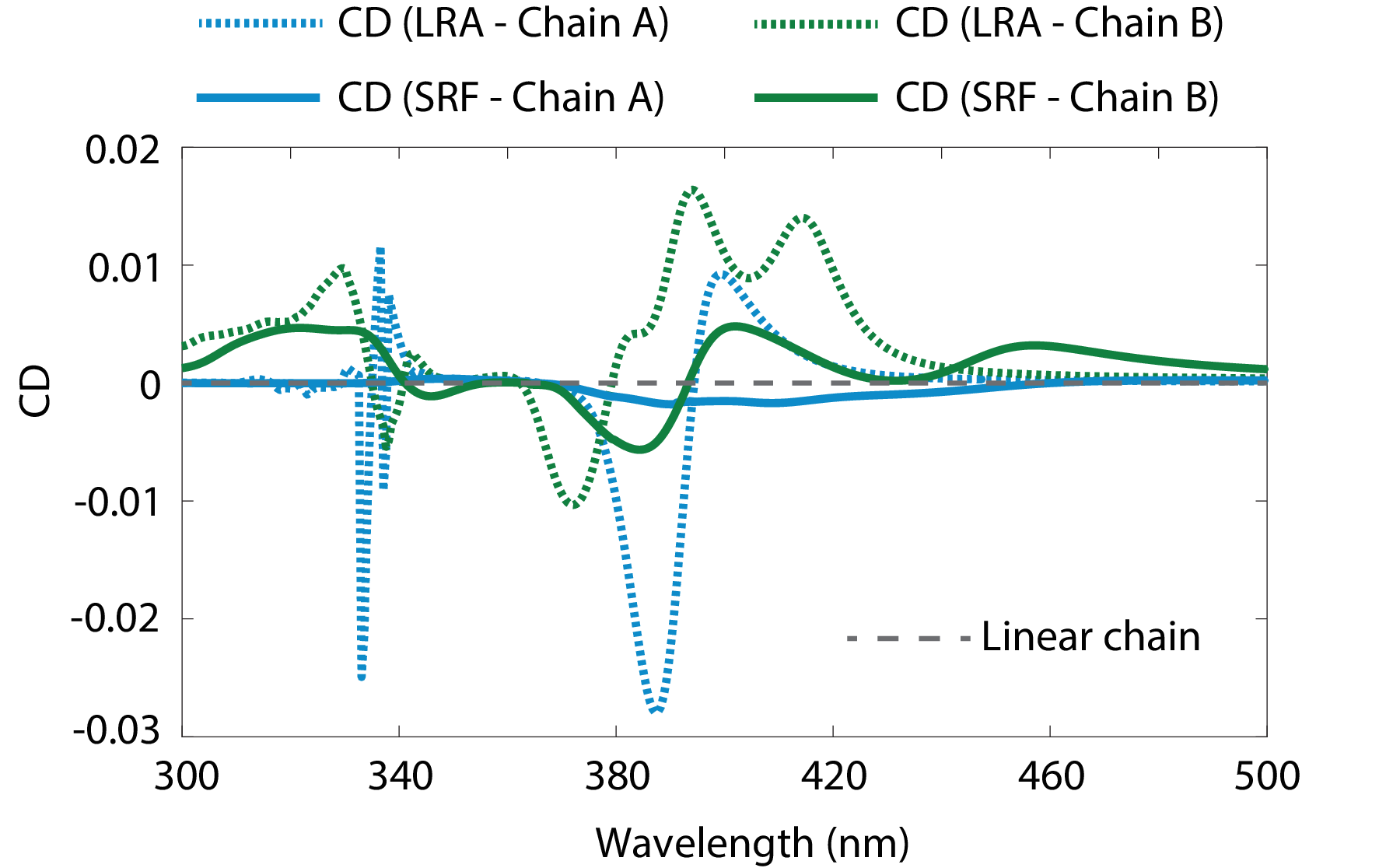}
\caption{
CD spectrum calculated within LRA (dotted lines) and SRF (solid lines)
for Chiral Chain A (green color) and Chiral Chain B (light-blue color).
The gray line represents the linear nanolens, where CD is obviously
absent.
}
\label{figure7}
\end{figure}

Another key point for the characterization of enantioselective
light--matter interactions is based on CD measurements,
as quantified by eq~\ref{Eq:CD}.
Fig.~\ref{figure7} contrasts Chiral Chains A and B (green and
light-blue color, respectively) with the linear chain (gray color),
in the cases of LRA (dotted lines) and SRF (solid lines). Obviously,
for an achiral geometry CD is non-existent and only manifests when
the symmetry of the linear chain is broken,
even for a planar-chiral case.
As soon as this occurs, both within LRA and when quantum corrections
are introduced, non-negligible CD emerges, as absorption in the system
is preferentially enhanced under polarization that is compatible with
its geometry. Moreover, in the framework of SRF, for both chains the
peaks are fewer, due to resonance broadening. We should highlight the
solid lines of Fig.~\ref{figure7}, which display that the CD spectrum
is fully affected by the consideration of quantum corrections, and for
a wide range of wavelengths it even takes opposite values compared to
the LRA case. Although the LRA CD spectrum gives impressive outcomes
for various nanophotonic environments, there is a strong possibility
that in reality the spectrum is completely different, as mesoscopic
phenomena have to be included in the case of NPs with radius below
$10$\,nm.

\begin{figure}[h!]
\centering
\includegraphics[width=0.9\columnwidth]{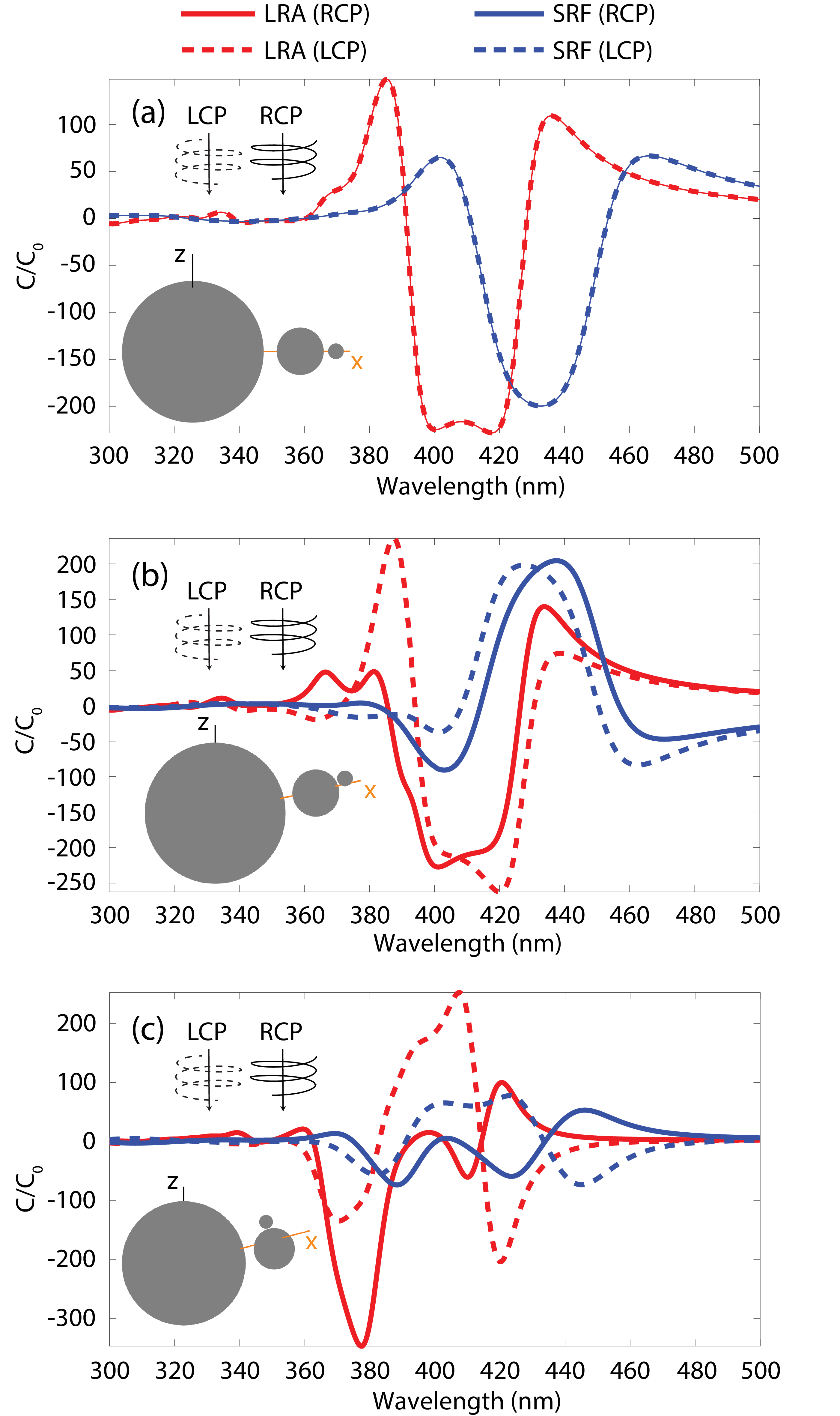}
\caption{
Normalized optical chirality $C/C_0$ at the $g_{2}$ hotspot in the case of
LRA (red color) and SRF (blue color), for RCP (solid lines) and LCP (dashed
lines), for (a) a linear chain, (b) Chiral Chain A, and (c) Chiral Chain B.
}
\label{figure8}
\end{figure}

Finally, in order to elucidate the profile of the optical chirality 
at the $g_{2}$ plasmonic hotspot and concurrently distinguish its
dependence on light polarization, we quantify in Fig.~\ref{figure8}
the optical chirality enhancement compared to vacuum ($C/C_0$). One
can notice that for an achiral structure, i.e., collinear NPs (see Fig.~\ref{figure8}a), the light polarization does not play a role,
yet the overall enhancement is quite important, as a result of the
scattered near fields. In contrast, for a
planar-chiral structure (see Fig.~\ref{figure8}b and Fig.~\ref{figure8}c)
the $C$ spectrum is significantly modified by the polarization of the
incident EM field, especially in the wavelength region of the chain
modes, which hints a completely different interaction of the fields
at the hotspot. It should be noted once again that the SRF treatment
reveals that LRA is insufficient to quantify the chiraloptical response
of the nanostructure. As seen in Fig.~\ref{figure8}, in the wavelength
window where the chirality density is maximized at the $g_{2}$ hotspot according
to LRA, SRF shows it to be minimized. Nevertheless, the enhancement of
$C$ is present in every scenario and we calculated enhancement factors
even higher than $300$ at about $380$\,nm in the case of Chiral Chain B,
as pointed out in Fig.~\ref{figure8}c.

In a nutshell, in the case of perfectly aligned nanolenses, the
chiroptical response is ascribed to light-induced chirality stemming
from the asymmetric coupling strength and phase differences between
the plasmonic modes around the NPs, whereas in the case of imperfect
nanolenses, it is ascribed to both planar chirality and interaction-induced
effects. Our study suggests that the plasmonic lensing effect can be
robust for other metallic nanochains as well, since we demonstrated
that nonlocality and mesoscopic phenomena do not considerably impair
the near-field enhancement. Regarding the chiroptical physics, the
main message of our work is that it is does not appear possible to
straightforwardly predict or quantify the chiroptical response of
plasmonic NPs of radii smaller than $\approx 10$\,nm without
considering nonlocality and quantum effects.

\section{Conclusions}
We theoretically investigated plasmonic nanolenses made of three Na
nanospheres with hierarchically decreasing radii and separations, with
the intention of exploring the influence of nonlocality on chain LSPs
and the resulting near-field intensity, with particular emphasis on
electron density spill-out and Landau damping as captured by SRF. We
demonstrated the efficiency of 
imperfect
self-similar plasmonic nanolenses through a statistical analysis of a
large number of possible chain arrangements, stemming from manufacturing
uncertainties, showing that the field enhancement remains robust, even
when the positions of the NPs deviate from the ideal straight line.
Subsequently, we concentrated on the chiroptical response of the perturbed
nanosystem. Quantum-corrected calculations suggest that the near-field
enhancement achievable in realistic configurations with broken symmetry can
lead to significant increase of optical chirality and circular dichroism.
We also showed that an increase of optical chirality density can be
accomplished regardless of the existence of a chiral geometry, provided
that there is both electric and magnetic field enhancement,
along with parallel orientation, spatial overlap, and an appropriate
phase difference.
While the increase of optical chirality density cannot be straightforwardly
predicted, both $C$ and CD can be manipulated through displacements of the
NPs, but there is no fundamental rule predicting which arrangement and
parameters will work better. Eventually, it is a combination of several
factors that determine the scattered fields, to create strong vortices of
fields around specific points in the vicinity of the structure.
Quantification of optical chirality is a complicated task, and measures
of chirality can often be misleading if nonlocality and quantum effects
are not taken into account.

\section*{Author contributions}

C. T. and N. K. conceived the idea, while N. A. M. promoted the statistical analysis. N. K. and X. Z. performed the calculations.
X. Z. developed the computation code. All authors contributed to the discussion of the results and writing the manuscript.

\section*{Acknowledgments}

The Center for Polariton-driven Light--Matter Interactions (POLIMA)
is sponsored by the Danish National Research Foundation
(Project No.~DNRF165).
X.~Z. was supported by the KU Leuven internal funds: the C1 Project
No. C14/19/083, the IDN Project No. IDN/20/014, and the small
infrastructure Grant No. KA/20/019.

\newpage

\bibliography{references} 

\end{document}